\documentclass[format=acmsmall, review=false, screen=false]{acmart}

\usepackage{graphicx}
\usepackage{textcomp}
\usepackage{xcolor}
\usepackage{import}
\usepackage{hyperref}
\usepackage{tabularx}
\usepackage{flushend}
\usepackage{multirow}
\usepackage{algorithm}
\usepackage[noend]{algpseudocode}
\usepackage{amsmath}
\usepackage{subcaption}
\usepackage{mathtools}
\usepackage{booktabs} 

\usepackage{tikz}

\newcommand{\redcircleletter}[1]{%
    \begin{tikzpicture}[baseline={(0,-0.5ex)}]
        \node[draw=red, circle, minimum size=1em, inner sep=1pt] {#1};
    \end{tikzpicture}%
}

\acmJournal{TECS}
\acmYear{2024} \acmVolume{1} \acmNumber{1} \acmArticle{1} \acmMonth{1}

\begin{document}

\title{Revealing CNN Architectures via Side-Channel Analysis in Dataflow-based Inference Accelerators}


\author{Hansika Weerasena}
\affiliation{%
  \institution{University of Florida}
  \city{Gainesville}
  \state{FL}
  \postcode{32611}
  \country{USA}
}
\email{hansikam.lokukat@ufl.edu}

\author{Prabhat Mishra}
\affiliation{%
  \institution{University of Florida}
  \city{Gainesville}
  \state{FL}
  \postcode{32611}
  \country{USA}
}
\email{prabhat@ufl.edu}

\begin{abstract}

Convolutional Neural Networks (CNNs) are widely used in various domains, including image recognition, medical diagnosis and autonomous driving. Recent advances in dataflow-based CNN accelerators have enabled CNN inference in resource-constrained edge devices. These dataflow accelerators utilize inherent data reuse of convolution layers to process CNN models efficiently. Concealing the architecture of CNN models is critical for privacy and security. This paper evaluates memory-based side-channel information to recover CNN architectures from dataflow-based CNN inference accelerators. The proposed attack exploits spatial and temporal data reuse of the dataflow mapping on CNN accelerators and architectural hints to recover the structure of CNN models. Experimental results demonstrate that our proposed side-channel attack can recover the structures of popular CNN models, namely Lenet, Alexnet, VGGnet16, and YOLOv2. 

\end{abstract}



\maketitle

\vspace{-0.1in}
\section{Introduction}
\label{sec:introduction}

Convolution Neural Networks (CNNs) \cite{gu2018recent} are Deep Neural Networks (DNNs) that incorporate convolution (Conv) layers specialized in processing multidimensional data. CNNs are used in a wide range of applications, such as image and video recognition, classification, and analysis. These neural networks operate in two main phases: training and inference. The training phase involves a time-consuming process of learning weights in the neural network, while inference uses the pre-trained neural network to perform fast predictions. 
Resource-constrained edge devices perform inference in the device rather than sending data to a centralized server for inference. These edge devices can range from mobile phones to remote offline sensor networks. 


Edge artificial intelligence (AI) enhances system performance by mitigating communication bottlenecks between edge devices and servers, ensuring high availability independently of network/internet connectivity, providing real-time insights, and minimizing the need for extensive data storage~\cite{sze2017efficient}. AI at edge needs inference of pre-trained CNN models on resource-constrained devices with energy and area constraints. General-purpose central processing units (CPUs) or graphics processing units (GPUs) can be used for the purpose of training and inference of CNNs, though GPUs are preferred for their superior parallel processing capabilities. Although central CPUs or GPUs are used in training these CNNs at servers, dataflow-based accelerators are preferred for inference at edge devices~\cite{chen2019eyeriss}. Dataflow is a computing scheme that utilizes inherent data reuse of Conv layers to achieve efficient CNN inference performance. Dataflow-based CNN accelerators save energy and execution time by reducing the cost of main memory accesses by introducing a local memory hierarchy inside the accelerator \cite{sze2017efficient}. Dataflows can be broadly classified into four categories depending on the type of stationary data~\cite{sze2017efficient}: weight stationary, output stationary, input stationary,
and row stationary. This paper explores two widely used dataflows: weight stationary (WS) and output stationary (OS). These dataflow accelerators can be designed using Application Specific Integrated Circuit (ASIC) as well as Field-Programmable Gate Array (FPGA). ASIC-based accelerators are preferred in edge inference since they are energy efficient while providing adequate computational flexibility~\cite{kwon2018maeri}. 


These CNN models need to be run on various edge devices with diverse hardware, software, and firmware developed by different vendors. Supply chain vulnerability can lead to security concerns for these edge devices with accelerators. For example, hardware Trojans can be inserted during the design as well as fabrication phase of an ASIC-based accelerator with the malicious intent of leaking sensitive information without being detected at the post-silicon verification stage or during runtime~\cite{mishra2017hardware}. In many application scenarios, the structure of a CNN model should be kept confidential for the following reasons. (1) CNN model can be a company's proprietary and critical intellectual property. (2) Knowing the network model leads to designing and launching efficient adversarial attacks~\cite{chakraborty2018adversarial}. (3) User privacy can be compromised in a shared accelerator if the model architecture is leaked, as it reveals unique data processing characteristics and intended uses of the model, while also aiding adversaries in executing model inversion attacks~\cite{fredrikson2015model}. Different types of side-channel analysis (memory, timing, electromagnetic emanation) are used in recovering CNN structures from GPU/CPUs \cite{hu2019neural,jha2020deeppeep,wei2020leaky,isakov2018preventing,yu2020deepem}. GPUs/CPUs use a temporal computing paradigm where centrally controlled processing units can only fetch data from the memory hierarchy. On the other hand, dataflow-based accelerators have a spatial computing paradigm where transfer between individually controlled processing units is possible. Due to the inherent difference in computing paradigm and underlying architecture, existing side-channel attacks on GPU/CPU-based accelerators cannot be directly applied to dataflow-based accelerators. Furthermore, existing memory-based side-channel attacks~\cite {hua2018reverse} on CNN processing focus on main memory to accelerator memory transfer, which leads to reverse-engineering a large set of possible CNN structures for a single CNN model. For example, ~\cite {hua2018reverse} gives 24 possible structures for Alexnet~\cite{krizhevsky2017imagenet}. 

In this paper, we try to answer a fundamental question: \textit{ is it possible for an adversary to exploit inherent data reuse of dataflow-based CNN inference accelerators via memory side-channels to accurately recover architectures of CNN models?} Our proposed research needs to answer two major challenges in developing such an attack: (1) how to exploit different dataflow patterns to converge a large number of potential structures to a few, although different layer structures can result in the same side-channel values, (2) how to develop a generalized approach to recover structures from different input, output, layer sizes and their mapping on the accelerator?  CNN consists of a sequence of Conv, fully connected (FC), and pooling layers. Due to the prevalence of Conv layers and their inherent data reuse, CNN accelerators focus on accelerating convolution layers. Therefore, our study principally concentrated on exploiting these acceleration techniques to recover CNN architectures. Specifically, this paper makes the following important contributions.

\begin{itemize}
    \item We adapt and refine the bus snooping-based threat model used in GPUs~\cite{hu2020deepsniffer, hua2018reverse} to collect memory-based side-channel information from dataflow-based CNN inference accelerators.
    \item We propose a mechanism to recover the structure of Conv and FC layers from weight stationary and output stationary dataflow-based CNN inference accelerators with local forwarding.
    \item Pooling layer parameters from the pooling module inside the CNN accelerator is recovered under assumptions on typical pooling layer characteristics.
    \item We propose a framework to recover the complete architecture of CNN by iterative recovery of individual layers.
    \item Experimental results demonstrate that our approach can fully recover CNN architectures from popular CNN models (Lenet~\cite{le1989handwritten}, Alexnet~\cite{krizhevsky2017imagenet}, VGG-16~\cite{simonyan2014very},  and YOLOv2~\cite{redmon2016you}).
\end{itemize}

The rest of the paper is organized as follows. Section \ref{sec:background}
presents background on CNN and dataflow-based accelerators. Section \ref{subsec:threat_model} outlines the threat model. Section \ref{sec:methodalogy} describes our proposed approaches for extracting CNN architectures. Section \ref{sec:experiments} presents the experimental results. Finally, Section~\ref{sec:conclusion} concludes the paper.

\section{Background and Related Work}
\label{sec:background}

This section first introduces DNNs and CNNs, focusing  on the convolution operation. It then outlines two primary architectures for processing CNNs:  temporal architecture and  spatial (dataflow-based) architecture. The discussion emphasizes dataflow-based architectures, which are the main focus of this paper. Finally, we review prior research on model extraction attacks on DNNs.

\subsection{Convolutional Neural Networks (CNN)}
\label{subsec:cnn_basics}

Drawing inspiration from human brain, neural networks simulate a behavior of neurons where each neuron computes a weighted sum of its inputs. This computation is not merely a linear operation; it involves a nonlinear function (activation function) that activates the neuron only if the weighted sum of inputs surpass a certain threshold. In a typical neural network, multiple layers are stacked, each consisting of numerous neurons. These neurons are interconnected across layers through connections known as weights, which determine the strength and influence of one neuron's output on the next layer's input. The input layer receives values and transmits them to subsequent layers, often including one or more `hidden layers'. These hidden layers process the weighted inputs and pass them on to the output layer, which then delivers the final results. DNNs involve architectures with multiple hidden layers, making these networks capable of serving as universal function approximators. Specifically, CNNs, a specialized type of deep neural networks, incorporate convolutional layers that excel at handling multi-dimensional data like images and videos.

\begin{figure}[htp]
\includegraphics[width=0.8\textwidth]{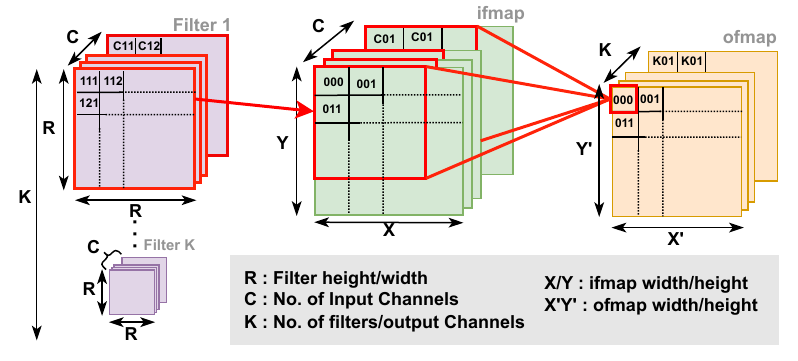}
\vspace{-0.1in}
\caption{Convolution layer parameters, semantics and operation: An $R \times R \times C$ filter is applied on same size neighborhood of {\it ifmap} of size $X \times Y \times C$ to calculate single value in {\it ofmap} of size  $X' \times Y' \times K$. }
\label{fig:computation_in_cnn}
\end{figure}

A typical input to a CNN is a matrix, such as an image, where each pixel value serves as an input. An input can have multiple channels, for example, if the image is in RGB format, it will have three channels, corresponding to the red, green, and blue components. CNN mainly consists of three types of layers: convolutional, pooling, and fully connected (FC)  layers \cite{gu2018recent}. Convolution layers dominate the computations in CNN (about 90\% \cite{cong2014minimizing}). A convolution layer takes an input activation/input feature map (\textit{ifmap}) and does 2-D convolution using a set of filters with weights to obtain an output feature map (\textit{ofmap}). Applying different filters results in extracting different embedded features from the \textit{ifmap}. Figure~\ref{fig:computation_in_cnn} illustrates a typical convolution operation with $K$ convolution filters of size $R$ and having $C$ input channels. It shows how the first element of \textit{ofmap} is calculated by the sum of element-wise multiplication between a filter of size $R\times R\times C$ and the same size neighborhood of \textit{ifmap}. Stride and padding are two other vital parameters of a convolution layer. Stride ($St$) represents the number of values a filter moves horizontally or vertically during a convolution operation. Padding ($Pd$) is the number of additional values around the edges of the \textit{ifmap}  before applying the convolution filter. All convolution layers follow the relationship in Equation~\ref{eq:st_pd}. The same relationship holds for $Y$ and $Y'$. 
\begin{equation} \label{eq:st_pd}
X' = ((X - R + 2Pd)/St)+1
\end{equation}

In an FC layer, all the values of \textit{ifmap} are connected to all the values of \textit{ofmap}. In other words, a single \textit{ofmap} value is composed using the weighted sum of all the \textit{ifmap} values. The pooling layer is typically used after Conv layer to reduce the dimensionality of the feature map. Max pooling and average pooling are two frequently used pooling operations~\cite{goodfellow2016deep}. Conv and FC layer execution can be viewed as a set of multiply and accumulate (MAC) operations, and modern accelerators perform a large number of MAC operations in parallel.

\subsection{Architectures for CNN Inference}
\label{subsec:cnn_architecture}

In the realm of deep learning, there are two critical phases: training and inference. The training phase involves learning to perform a specific task, such as classifying an image into a designated label. This process primarily focuses on determining the optimal weights within the network using a training dataset. Specifically, within the convolutional layers, this phase is responsible for learning the filter weights. After training, the network used in the inference phase, using predetermined weights to calculate outputs for new input data and perform its designated task. This paper specifically examines the inference phase and the hardware architectures that facilitate it, aiming to explore model extraction attacks.

Due to the growing popularity of DNNs, specialized hardware features have been developed to optimize DNN processing. Highly-parallel compute architectures are instrumental in achieving optimal performance during the inference phase, often paralleling MAC operations. These architectures can be classified into two paradigms: temporal and spatial~\cite{sze2017efficient}. Temporal architectures, commonly found in CPUs and GPUs, enhance parallelism through vectors (SIMD) or parallel threads (SIMT) and maintain a centralized control over numerous Arithmetic Logic Units (ALUs). These ALUs are restricted to fetching data from the memory hierarchy without direct inter-ALU communication. They map both fully connected (FC) and convolutional (Conv) layers to matrix multiplications for parallelized MAC operations. For instance, the convolution operation can be reformulated as matrix multiplication by converting one of the inputs into a Toeplitz matrix~\cite{strang1986proposal}, which is used in temporal accelerators. In spatial (dataflow) architecture, each ALU is equipped with its own control logic and local memory. An ALU with its own local memory is referred to as a processing element (PE). These PEs form a processing chain allowing direct data transfer between them. These architectures do not necessitate mapping FC and Conv layers to generic matrix operations, enabling a more natural execution of DNN layers via individual MAC operations. Dataflow accelerators mitigate the memory bottlenecks common in temporal architectures by introducing multiple levels of local memory hierarchy and interconnects. These include global buffers connecting to DRAM, small local registers within each PE, and an inter-PE network facilitating direct data transfers between PEs. This memory structure significantly boosts energy efficiency by reducing the cost of data accesses by fetching data from local registers or neighboring PEs consuming substantially less energy than accessing DRAM~\cite{sze2017efficient}. Due to the differences in architecture and processing between the two paradigms, existing attacks designed for model extraction on CPUs/GPUs cannot be directly applied to dataflow architectures.

\color{black}

\subsection{Dataflow-based CNN Accelerators}
\label{subsec:dataflow-based DNN accelerators}





The dataflow-based CNN inference accelerators considered in this study process the CNN layer by layer~\cite{hua2018reverse}. In other words, the accelerator loads {\it ifmap}/weights of a particular layer to the global buffer, processes it, and writes {\it ofmap} of the layer back to the main memory. Dataflow determines how the data is moved and processed through the accelerator architecture to perform MAC operation needed for a CNN layer. Dataflow mapping refers to how MAC operations are assigned to each processing element in each cycle.   Dataflow-based CNN accelerators try to get maximum data reuse efficient dataflow mapping. A typical architecture~\cite{machupalli2022review,hua2018reverse} of a dataflow-based CNN inference accelerator used in edge devices is shown in Figure~\ref{fig:overview_of_accel}. The controller, interconnects, global buffer (GB), and PE arrays can be identified as critical components of a CNN accelerator. 

A typical CNN accelerator supports three separate Network-on-Chip (NoC)/interconnects ~\cite{kwon2018maeri,chen2019eyeriss} for two reasons: (1) different data types (weights/\textit{ifmaps}/{\textit{ofmaps}}) need different data transmission patterns (unicast, multicast, and broadcast), and (2) enables high-speed data transfer and pipelined operation. Dataflow defines how PEs and interconnects are arranged. While there are several types of dataflows, including weight stationary (WS), output stationary (OS), input stationary (IS), and row stationary (RS), this paper focuses on two most widely used configurations: WS and OS. There are WS~\cite{gokhale2014240,chakradhar2010dynamically,kwon2018maeri} and OS~\cite{du2015shidiannao,moons201714} dataflow based architectures with subtle differences. Our study focuses on the dataflow aspect of the accelerator by using a generic WS (Section \ref{subsec:ws_dataflow}) and OS (Section \ref{subsec:os_dataflow}) dataflow architecture with input forwarding. By proposing an attack on these generic dataflow models, we enhance the generalizability and adaptability of our findings, making them applicable to a broad range of accelerators.

\subsubsection{Weight Stationary Dataflow}
\label{subsec:ws_dataflow}

\begin{figure}[htp]
\includegraphics[width=0.8\textwidth]{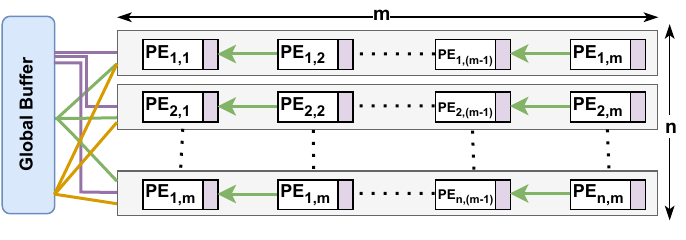}
\vspace{-0.2in}
\caption{\textit{WS}($m,n$) architecture: weight stationary dataflow with \textit{n} processing element (PE) arrays and \textit{m} PEs per array, with separate interconnects for weight/input reads and partial sum reads and writes from/to Global Buffer (GB). Neighboring PEs in an array has input-forwarding connections.}
\label{fig:ws_generic_asic}
\vspace{-0.1in}
\end{figure}

\begin{figure}[tp]
\centering
\includegraphics[width=0.7\textwidth]{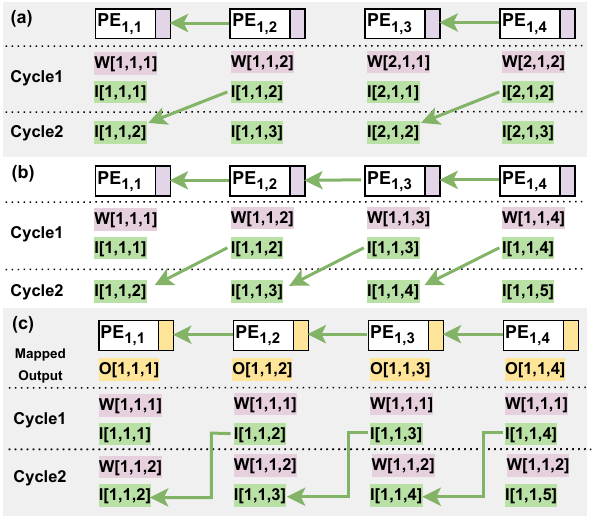}
\vspace{-0.1in}
\caption{Dataflow mapping and data reads for the first two cycles for three scenarios: (a) mapping of $2 \times 2 \times 2$ filter into \textit{WS}($4,1$),  (b) mapping of $4 \times 4 \times 2$ filter into \textit{WS}($4,1$), and (c) mapping of $2\time 2\times 2$ filter into \textit{OS}($4,1$) with $st=1$ and $pd=0$ and {\it ifmap} of $\{X=5, Y=5, C=2\}$, and {\it ofmap} of $\{X'=4, Y'=4, K=1\}$.}
\label{fig:dataflow_mapping_ws_and_os}
\end{figure}

As the name suggests, once a weight value is read to a PE, a WS dataflow does all the MAC operations involving that weight before reading a new value to a PE. Figure \ref{fig:ws_generic_asic} shows an architecture of a weight stationary dataflow that reuses weights temporarily and input activations spatially. It has $n$ PE arrays, each with $m$ PEs and denoted by the notation \textit{WS}($m,n$). This accelerator has unicast NoC for weight, PE array-wise multicast supported NoC for inputs, another interconnect for partial sum reads, and a fixed accumulation tree NoC with adders similar to~\cite{kwon2018maeri}. There are separate adder trees per array, and an adder tree will accumulate/sum up all weight activation products of an array in the preceding cycle. The resulting value is a partial sum (\textit{psum}) for a one {\it ofmap} entry. This architecture can process multiple filters simultaneously using separate PE arrays per filter. This architecture minimizes {\it ifmap} reads by spatial reuse in two ways: (1) PE array-wise multicast shares the same input values across multiple PE arrays where each PE array calculates for a different filter, and (2) forwarding connection in a PE array can share activation values between two cycles occurring due to the stride of the filter. This weight stationary architecture has flexibility in mapping, depending on the layer, the in-between forwarding connections between PEs can be switched on/off. Mapping of the layer to the PEs is optimized for maximum PE utilization and minimizing input reads. For example, if $m = 4$ and $n=1$, the first row of two channels of $2 \times 2 \times 2$  CNN filter can be mapped as shown in Figure \ref{fig:dataflow_mapping_ws_and_os}(a). If the filter size is $4 \times 4 \times 2$, only one channel row of the filter is mapped in the first cycle as shown in Figure \ref{fig:dataflow_mapping_ws_and_os}(b). Both of the examples have a stride of one. State-of-the art MAERI accelerator architecture~\cite{kwon2018maeri} employs the basic WS dataflow detailed in this section. 

\subsubsection{Output Stationary Dataflow}
\label{subsec:os_dataflow}

An output-stationary dataflow accumulates \textit{psums} corresponding to one {\it ofmap} value in the internal register of a PE until it is fully calculated. In other words, a PE is mapped to a single value of {\it ofmap} until that value is fully calculated. Similar to the WS dataflow architecture discussed in Figure~\ref{fig:ws_generic_asic}, it has $n$ PE arrays, each with $m$ PEs and denoted by the notation \textit{OS}($m,n$), but with the following modifications. Instead of the internal register keeping weights stationary, it accumulates \textit{psums}. This accelerator has unicast NoC for inputs, single value broadcast supported NoC for weights, and separate interconnect to write outputs to the GB. One PE array will accumulate sums relevant to one row of {\it ofmap} so that forwarding links between PEs in the array can have maximum utilization. Due to the broadcast of a single weight, every PE calculates MAC relevant to one input channel in a cycle. In the first cycle, the accelerator multicasts the same weight to all the PEs and unicasts relevant activations to each PE. After doing MAC operations and accumulating the partial sum to the internal registry, different weight is broadcast in the second cycle. After the local forwarding of inputs, the remaining  inputs are unicast relevant to the previous partial sum. Figure~\ref{fig:dataflow_mapping_ws_and_os}(c) shows a layer mapping with one filter with parameters $\{R=2, C=2, st=1, pd=0\}$, {\it ifmap} of $\{X=5, Y=5, C=2\}$, and {\it ofmap} of $\{X'=4, Y'=4, K=1\}$ to a $m=4$ and $n=1$ output stationary accelerator. As we can see, there is only one input read (I[1,1,5]) in the second cycle due to spatio-temporal forwarding of inputs (I[1,1,2] and I[1,1,3]) from the previous cycle. State-of-the art Shidiannao accelerator architecture~\cite{du2015shidiannao} employs the basic OS dataflow detailed in this section. 

\subsection{Related Work}
\label{subsec:related_work}

There are many efforts on leaking DNN/CNN architectures using side-channel attacks. Timing and memory side channels have been used to recover DNN models in \cite{wei2020leaky,hua2018reverse,hu2019neural,jha2020deeppeep}.  An attack to recover compact DNN models from GPU using timing, memory, power, and kernel side channels is proposed in \cite{jha2020deeppeep}. They assume the attacker knows power consumption, memory footprint, and latency for backward and forward propagation for different batch sizes. Hu et al. \cite{hu2019neural} propose a method to find DNN architecture by eavesdropping into off-chip data transfer between CPU and GPU and exploiting the entire DNN execution stack (the DNN library, Hardware abstraction, and Hardware). Another side-channel attack on analyzing main memory to accelerator memory access trace and execution time is discussed in \cite{hua2018reverse}. The adversary can control the input to the accelerator and derive a set of potential models for the currently running DNN model. For example, they propose 24 possible structures for Alexnet. Wei et al.~\cite{wei2020leaky} exploit context switching of GPU to recover DNN models. This attack is made in the training stage because they can exploit multiple uses of the same layers over training time. Similar threat model was proposed in~\cite{hu2020deepsniffer} that recovers DNN architecture through the acquisition of memory access events from bus snooping inside GPU during inference time. Apart from memory and timing side channels, attacks are proposed on cache side channels to recover DNN models \cite{isakov2018preventing, yan2020cache, liu2020ganred}. Several studies have demonstrated attacks using power~\cite{xiang2020open} and electromagnetic side channels~\cite{yu2020deepem, batina2019csi,horvath2023barracuda, horvath2024cnn,chmielewski2021reverse} to successfully recover CNN models form GPU and other temporal architectures. These studies validate that recovering DNN/CNN architecture is a critical security concern. All existing side-channel attacks to recover DNN/CNN model architectures focus on temporal inference accelerators such as GPUs. Due to inherent differences in architecture and processing mechanism, side-channel attacks on temporal architectures cannot be directly applied to dataflow-based inference accelerators. To the best of our
knowledge, our paper is the first study on recovering CNN model architecture using memory side channels of dataflow-based CNN inference accelerators.

\section{Threat Model and Problem Formulation}
\label{subsec:threat_model}


In this section, we define the problem space and establish the threat model under which our attack operates. This formulation is critical as it lays the groundwork for understanding the potential vulnerabilities within dataflow-based CNN accelerators.

\subsection{Problem Formulation}
\label{subsec:problem_definition}

\begin{figure}[tp]
\includegraphics[width=0.8\textwidth]{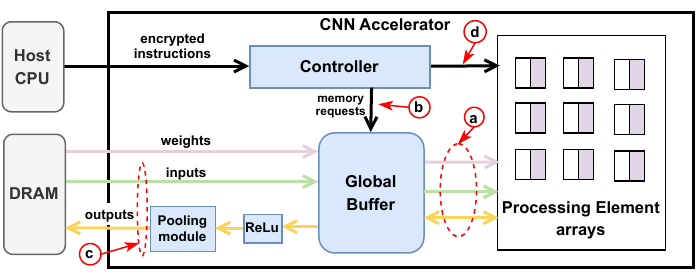}
\vspace{-0.10in}
\caption{Overview of a typical CNN accelerator. The controller, interconnects, global buffer (GB), and processing element (PE) arrays are the critical components. Unprotected communication can be snooped to extract side-channel data. }
\label{fig:overview_of_accel}
\end{figure}

Figure \ref{fig:overview_of_accel} illustrates an overview of a typical dataflow-based CNN accelerator with layer-by-layer execution described in Section~\ref{subsec:dataflow-based DNN accelerators}. If we zoom into the processing of one layer, first, the host CPU loads {\it ifmap}/weights to the GB. These \textit{ifmaps}/weights are stored as arrays in memory, which means they are stored in contiguous memory locations, and GB preserves the order. The PE arrays are designed in a pipelined manner to do MAC operations in each cycle. Each PE in the PE array has a local registry to hold or accumulate certain data elements to reuse in MAC operations. The controller is responsible for loading weights/inputs/\textit{psums} to individual PEs in each cycle. The controller first calculates memory requests for the next cycle depending on the dataflow supported by the accelerator and layer parameters. Then, it issues memory requests to the global buffer. The global buffer uses interconnects between GB and PE arrays for data transfer. The focus of this paper is to extract CNN model architecture based on memory side-channel information from dataflow-based CNN accelerator.

\subsection{Threat Model}
\label{subsec:threat_model_details}

 The threat model assumes the adversary gathers the following memory side-channel information regarding the execution of each layer CNN: (1) the total number of weight/input reads and output writes, (2) the number of weight/input reads, output writes each cycle of execution until a \textit{targeted event} ($e$) that depends on the dataflow (the concrete definition of $e$ in WS and OS dataflow are stated in Section~\ref{subsec:recover_conv_ws} and ~\ref{subsec:recover_conv_os} respectively.),  and (3) output stationary dataflow needs the virtual address of weight reads in the first two cycles. Here, an input read means reading of a single value from an \textit{ifmap} and an output write means writing of a single value to an \textit{ofmap}. Depending on the adversary's capabilities, there are several ways to obtain these three pieces of information. The adversary can exploit unprotected communication between accelerator components. Information (1) and (2) can be obtained by snooping interconnects from GB to PE arrays (\redcircleletter{a}
 in Figure \ref{fig:overview_of_accel}). Since these accelerators use separate interconnects for each data type, it is easy to distinguish and count  GB accesses that can be snooping on the interconnects. Alternatively, the adversary can obtain information (1) and (2) along with (3) by snooping the unprotected bus between the controller to GB (\redcircleletter{b} in Figure \ref{fig:overview_of_accel}). If there is a pooling layer, for the recovery pooling layer parameters, the adversary needs to find the number of pooling operations ($N_{pool}$).  $N_{pool}$ can be recovered using the number of output writes to DRAM from GB (\redcircleletter{c} in Figure \ref{fig:overview_of_accel}) during the layer execution. 
 
Bus snooping is a low-cost, practical and well understood attack that has been widely demonstrated~\cite{huang2002keeping,huang2014hmtt,blass2012tresor}. Snooping of these buses (\redcircleletter{a}, \redcircleletter{b} and \redcircleletter{c}) can be done either through lightweight Hardware Trojans (HTs) or physical microprobing of buses. The reliance on a long, distributed, and potentially untrusted supply chain in chip design raises the risk of malicious implants, such as HTs, introduced through various channels including untrusted CAD tools, rogue designers, or directly at the foundry \cite{mishra2017hardware}. An HT can be clandestinely inserted into the RTL or netlist of an accelerator and remain undetected not only at the post-silicon verification stage but also during runtime~\cite{mishra2017hardware}. Furthermore, simple multiple HTs, such as HTs with counters inserted in on-chip communication for snooping are harder to detect in runtime due to their insignificant area and power overhead~\cite{ahmed2020defense, weerasena2023security}. Advances in technology have significantly improved microprobing capabilities, enabling the extraction of encryption keys, intellectual property, and personal data from densely packed hardware implementations, as demonstrated in recent studies~\cite{shi2016layout,skorobogatov2017microprobing}. Notably, optical probing has been used to obtain FPGA bitstream decryption keys from 28nm Xilinx devices~\cite{tajik2017power}, and focused ion beam techniques have effectively extracted data from SoC buses~\cite{wang2017probing}. These methods illustrate that precise data extraction from buses can be leveraged in our attack to gather side-channel information. Alternatively, compromised accelerator firmware or host operating systems could also exploit performance counters to stealthily extract relevant side-channel information ((1), (2), and $N_{pool}$), complementing the physical snooping methods described earlier.

Our threat model also assumes that the adversary knows the accelerator architecture (number of PE arrays, number of PEs in an array, and the bandwidth of interconnects for each data type) as well as the dataflow mapping of each accelerator described in Section~\ref{subsec:dataflow-based DNN accelerators}. Typically, these basic architecture details are readily accessible for open-source or standard data-flow accelerators~\cite{du2015shidiannao,moons201714,chen2016eyeriss}. We also assume the adversary knows the \textit{ifmap} parameters ($X,Y,C$) of the first layer.  Similar assumption is made in ~\cite{hua2018reverse} because adversary can control inputs to the accelerator. In the layer-by-layer recovery of CNN, knowing the previous layer \textit{ofmap} parameters can be directly used as \textit{ifmap} parameters for the next layer. In other words, knowing the \textit{ifmap} parameters of the first layer and recovering the parameters of that layer indirectly finds the \textit{ifmap} parameters of the second layer. The proposed attack is passive, where the adversary can only snoop on the side-channel information but cannot alter dataflow mapping or data communication inside the accelerator. Furthermore, the adversary does not need to know the training or testing data of the CNN model. 

The recovery of a CNN model’s architecture poses significant risks for the model owners and users. (1) If the structure of a CNN is disclosed, it becomes easier for adversaries to craft efficient adversarial attacks~\cite{chakraborty2018adversarial, hu2020deepsniffer} fine-tuned to mislead the model without detection, thereby facilitating evasion attacks. For example, if a particular model is used to detect and flag network anomalies, an adversary can manipulate network traffic in such a way that it evades detection, allowing malicious activities to proceed undetected and model owner failing to provide the intended functionality. (2) The exposure of the model architecture can lead to the leakage of proprietary and critical intellectual property. For many companies, the design of their models embodies valuable research and development efforts, and unauthorized access to this information can lead to competitive disadvantages and economic losses. (3) As outlined in our paper, the leakage of model architecture in environments where the hardware is shared, increases the risk of user privacy breaches. Adversaries can exploit the known architecture to perform more effective model inversion attacks~\cite{fredrikson2015model}, potentially reconstructing sensitive training data. In summary, there are various potential threats from the recovery of a models’s architecture, including feasibility of adversarial attacks, loss of intellectual property, and breaches of user privacy, all of which can have severe consequences for the model stakeholders.

\section{Extracting CNN Architecture using Side-Channel Analysis}
\label{sec:methodalogy}

Algorithm \ref{alg:overview} shows an overview of the attack for recovering potential CNN structures from a dataflow-based accelerator. Lines 3-9 elaborate layer-by-layer side-channel data collection, which needs recognizing layer boundaries (line 5) discussed in Section~\ref{subsec:identify_layer_boundary}. In each side-channel variable, the Superscript ($j$ or $k$) specifies the layer number relevant to the variable. Lines 10-14 elaborate layer-by-layer structure recovery. The first dimension of the $layers$ 2D array (line 10) contains the number of layers ($j$) in CNN, and the second dimension contains possible structures for each layer. In layer-wise structure recovery, as the first step, the layer type is identified (line 12), which is discussed in Section~\ref{subsec:identify_layer_type}. Then the side-channel information is used to recover each layer's parameters (line 14). It is important to notice that if there are multiple potential structures for the previous layer, the algorithm tries \textit{ifmap} parameters ($X,Y,C$) of all of them to recover potential structures for the current layer (loop at line 13). Finally, the adversary flatten the 2D array $layers$ to multiple 1D arrays satisfying $ifmap_j = ofmap_{j-1}$ to get all the potential structures of the CNN model (line 15).


\begin{table}[tp]
\caption{Table of notations.}
\vspace{-0.1in}
\begin{center}
\resizebox{0.60\textwidth}{!}{%
\begin{tabular}{r l }
\toprule
$n$ & Number of PE arrays in an accelerator \\
$m$ & Number of PEs per array in an accelerator \\
$W_r$ & Total no. of weight reads for the layer execution. \\
$I_r$ & Total no. of input reads for the layer execution. \\
$O_w$ & Total no. of output reads for the layer execution. \\
$t_e$ & Cycle number of the targeted event $e$. \\
$w$ & Array of no. of weight reads at each cycle. \\
$i$ & Array of no. of input reads at each cycle. \\
$o$ & Array of no. of output writes at each cycle. \\
$i[t]$ & Number of input reads at $t^{th}$ cycle. \\
$w[t]$ & Number of weight reads at $t^{th}$ cycle. \\
$\&(dt)$ & Virtual address of data $dt$ in global buffer. \\
$H$ & Possible parameter sets for a Conv Layer. \\
$ceil(x)$ & Round-up number x to nearest integer. \\
\bottomrule
\end{tabular}
}
\end{center}
\label{tab:tableOfNotation}
\end{table}

\begin{algorithm}
\caption{Recovering potential CNN structures }
\label{alg:overview} 
\begin{algorithmic}[1]
\State {\bf Input:} First {\it ifmap} parametes, \{$X_1,Y_1,C_1$\}
\State {\bf Output:} Potential CNN architectures
\State j = 1 \Comment{to count the number of layers}
\While{\textit{processing of the CNN}}
\While{\textit{IdentifyLayerBoundary()}}
    \State $W_{(r,j)},I_{(r,j)},O_{(w,j)} \gets$ collect total R/W counts
    \State $w_j,i_j,o_j \gets$ collect cycle-wise R/W counts until event $e$ 
    \State $N_{(pool,j)} \gets$ collect No. of pooling operations 
\EndWhile
\State j ++
\EndWhile
\State $layers = [ ]$ \Comment{empty 2D array}
\For{k=1 to j}
\State type $\leftarrow$ \textit{IdentifyLayerType($W_{(r,k)},I_{(r,k)},O_{(w,k)}, N_{(pool,k)}$)}
\For{number of layer structures in layers[k-1]}
\State $layers[k]$ $\cup$ \textit{recoverLayer}($W_{(r,k)}$,$I_{(w,k)}$,$O_{(w,k)}$,$w_k$,$i_k$,$o_k$, $X_k$,$Y_k$,$C_k$,$type$)
\EndFor
\EndFor
\State {\bf Return} Flatten 2D array $layers$ to multiple 1D arrays satisfying $ifmap_j = ofmap_{j-1}$ to get all potential structures.
\end{algorithmic}
\end{algorithm}

\begin{algorithm}
\caption{Recovering layer parameters }
\label{alg:overview_layer_rec} 
\begin{algorithmic}[1]
\Function{\textit{recoverLayer}}{$W_r,I_r,O_w,w,i,o,X,Y,C,type$}
\If {type = FC}
\State \textit{layer} $\leftarrow$ \textit{recoverFC}($W_r,I_r, O_w$)
\EndIf
\If {type = Conv}
\State \textit{H} $\leftarrow$ \textit{$recoverConv_{DF}$}($W_r, O_w,w,i,o,X,Y,C$)
\If {Conv layer has pooling}
\For{h in H}
\State  \textit{h} $\cup $ \textit{recoverPooling}($N_{pool}, X',Y',K$)
\EndFor
\EndIf
\State layer $\gets H $
\EndIf
\State \textbf{retrun} layer
\EndFunction
\end{algorithmic}
\end{algorithm}

\begin{figure*}[htp]
\centering
\includegraphics[width=\textwidth]{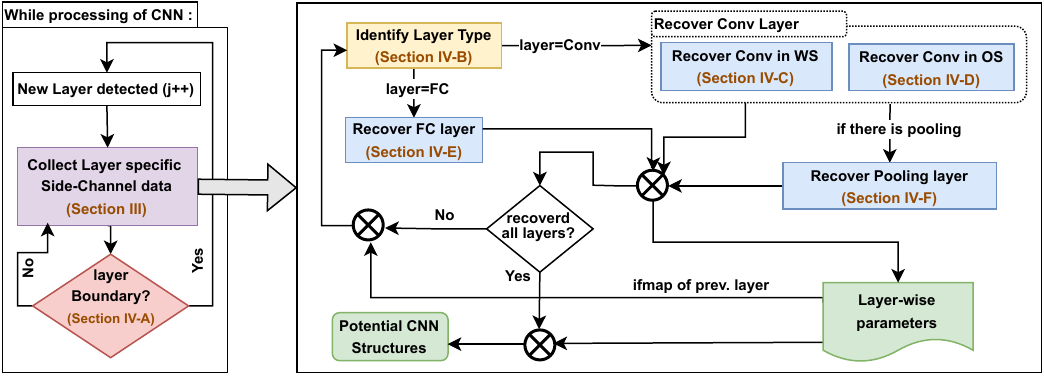}
\vspace{-0.1in}
\caption{Overview of the proposed side-channel attack to recover CNN model. The adversary collects layer-by-layer side-channel information, and utilizes them to recover layer-by-layer CNN structure.}
\label{fig:overview_of_algo}
\end{figure*}

Algorithm \ref{alg:overview_layer_rec} zooms into recovering individual layer structures. Depending on the layer type, the layer recovery procedure calls different functions (line 3, 5 and 8). The most crucial and difficult task is to recover Conv layer. The subscript in $recoverConv_{DF}$ highlights that the recovery of the Conv layer depends on the dataflow. Therefore, recovery of Conv from WS dataflow and OS dataflow are elaborated in Section \ref{subsec:recover_conv_ws} and \ref{subsec:recover_conv_os}, respectively. The function $recoverConv_{DF}$ returns a list of potential structures ($H$) but the conditional filters used in recovering Conv 
layers in WS and OS dataflows ensure that $H$ is a list with a few solutions or one solution (Section \ref{sec:experiments} shows Conv layer recovery in popular benchmarks converging to one structure). The recovery of FC (line~3) and pooling (line 8) layers are discussed in Section \ref{subsubsec:extarcting_fc_layer} and \ref{subsubsec:extractin_poool_layer}, respectively. Table~\ref{tab:tableOfNotation} outlines the notations used in these algorithms. Figure \ref{fig:overview_of_algo} presents a high-level overview of our proposed side-channel attack.

\subsection{Identification of Layer Boundary}
\label{subsec:identify_layer_boundary}

We present two ways to identify layer boundaries in a dataflow-based accelerator. The first method is to use the Read-After-Write (RAW) dependency of \textit{ifmap} current layer and \textit{ofmap} of the previous layer. Dataflow accelerators have sequential execution of the layers. Therefore, upon a successful layer execution, the \textit{ofmap} is written back to the hosts' main memory (DRAM) and read it back to GB in the next layer as \textit{ifmap}. This event can be identified using the memory trace as a RAW dependency on the same memory address. This needs an adversary needs to collect or actively observe the memory trace (address and cycle) between the GB and DRAM of the host. Similar to collecting previous side-channel information, this can be done by snooping the memory bus between GB to DRAM (\redcircleletter{c} in Figure \ref{fig:overview_of_accel}).

The second method is by identifying the configuration phase between layers. However, this approach is applicable in accelerators that have some reconfigurability. Since each layer differs, there is a configuration phase to dataflow mapping at each layer. Most CNN accelerators use a separate low-bandwidth bus (\redcircleletter{d} in Figure \ref{fig:overview_of_accel}) to send these configurations~\cite{kwon2018maeri}. Therefore, looking at active periods in that network is a simple way to identify boundaries. For example, the WS dataflow-based accelerator configures (on/off) forward links between PEs configured at each layer's beginning. The adversary can listen to this control message to identify layer boundaries.

\subsection{Identification of Layer Type}
\label{subsec:identify_layer_type}

The adversary needs to separate between FC, Conv, and Pooling layers. Because of no data reusability in FC layers, the total number of weight reads ($W_r$) in a layer is equal to the multiplication between the total number of input reads ($I_r$) and output writes ($O_w$). This relationship is used to distinguish an FC layer from a Conv layer. If there is a pooling layer after the execution of Conv layer, CNN accelerators run the Conv layer together with the pooling layer using a separate pooling module as shown in Figure~\ref{fig:overview_of_accel} (not processed as two separate layers). An adversary can detect a pooling layer by observing a difference between the number of outputs written to the global buffer and those written to DRAM inside the layer boundary. 















\subsection{Recovery of Conv Layer from Weight Stationary Dataflow}
\label{subsec:recover_conv_ws}

There are five parameters ($R,C,K,St,Pd$) that define a Conv layer. Since the number of input channels ($C$) is known from \textit{ifmap} parameters, an adversary needs to find four parameters ($R,K,St,Pd$) for successful Conv layer recovery. Algorithm \ref{alg:ws_attck_1} describes the attack steps for recovering a Conv layer from the WS dataflow based architecture outlined in Section \ref{subsec:dataflow-based DNN accelerators}. We refer to the $1^{st}$ cycle relative to each layer which is the first cycle with data transfer between GB to PE arrays. Here we define the \textit{targeted event} for the cycle-to-cycle data collection as the cycle $t_e$ where $i[1] = i[t_e]$. In other words, the adversary collects data reads/writes between the first cycle and the next ones with the same number of reads in the first cycle. This collects the number of data reads/writes accountable for the one row of the output feature map.

\begin{algorithm}
\caption{Recovering Conv Layer from WS dataflow}
\label{alg:ws_attck_1} 
\label{CnW:alg:Bit_by_bit_chaffing}
\begin{algorithmic}[1] 
\Function{\textit{$recoverConv_{ws}$}}{$W_r,O_w,w,i,o,X,Y,C$} 
\State $H \leftarrow$ solve Equation 
\ref{ws_w_read} $\in \mathbb{Z}^+$
\label{alg-ln:ws_attck_1_1} 
\State $n_a \leftarrow w[1]/i[1]$ 
\State $X' \leftarrow t_e - 1$
\label{alg-ln:ws_attck_1_3} 
\For{$\{R,K\}$ in H}
\If {$ R \leq w[1]/n_a$ and $(w[1]/n_a)\%R \neq 0$}
\State remove $\{R,K\}$
\EndIf
\State $st \leftarrow i[2]/max((m//R),1)$
\If {$st \notin \mathbb{Z}^+$ or $st > R$} 
\State remove $\{R,K\}$
\EndIf
\State $pd \leftarrow$ substitute $\{R,st\}$ and $X,X'$ to Equation~\ref{eq:st_pd} 
\If {$pd \notin \mathbb{Z}^+$ or $pd > R$ } 
\State remove $\{R,K,st\}$
\EndIf
\State $Y' \leftarrow$ substitute $\{R,st\}$ and $Y,pd$ to Equation~\ref{eq:st_pd} 
\If {$O_w \neq X' \times Y' \times K$} 
\State remove $\{R,K,st,pd\}$
\EndIf
\EndFor
\State \textbf{return} $H$
\EndFunction
\end{algorithmic}
\end{algorithm}

Due to the inherent feature of a WS dataflow, a single weight value is read only once: 
\begin{equation} \label{ws_w_read}
 W_{r} = R^2 \times C \times K
\end{equation}





Solving of Equation \ref{ws_w_read} with known $C$ in positive integer ($\mathbb{Z}^+$) domain provides potential solutions ($H$) for \{R,K\} values of the layer parameters (line \ref{alg-ln:ws_attck_1_1}). Then the adversary calculates $n_{a}$, the number of active PE arrays in the accelerator (line 3). While most of the time, the number of filters (\( K \)) is greater than the number of PE arrays (\( n \)) of the accelerator architecture, resulting in \( n_a = n \), the value of \( n_a \) is calculated as $w[1]/i[1]$ regardless of whether \( K \) exceeds \( n \) or not. Due to the \textit{targeted event} of data collection, the number of cycles between reveals the width of the output feature map ($X'$) (line \ref{alg-ln:ws_attck_1_3}). Even when $R > m$ and folding is supported, the number of cycles up to the event minus one reflects the value of $X'$. Then the adversary iterates (lines 5 - 16) through the potential \{$R,K$\} values and uses architectural hints and side-channel information to filter out incompatible \{$R,K$\}. The first condition (line 6) applies when the potential value of $R$ is smaller than the active PE array length ($w[1]/n_a$). As discussed in Section \ref{subsec:dataflow-based DNN accelerators}, the adversary uses the dataflow-mapping property of the accelerator to minimize input reads by only fully mapping filter rows. For example, if the PE array length is 12 ($m=12$) and $R, K$ are 5 and 4, the layer is mapped as two 10 MACs across two channels by utilizing only 10 PEs, not 12 Macs followed by 8. Line 6 checks the condition ($(w[1]/n_a)\%R \neq 0$) and filter out solutions. 

The adversary calculates the stride ($st$) of the respective potential $R$ (line 8). Since the forward links between PEs are used to forward \textit{ifmap} values from the previous cycle, the number of new input reads for the current cycle is always an integer multiple of stride. At line 8, $m//R$ (// is the integer division) gives out the number of channels of a filter mapped to the PE array when $m \geq R$. If $m<R$ and folding is used, only one channel is mapped to a PE array. Taking the maximum, consider both conditions ($m \geq R$ and $m<R$ ). The second cycle has input forwarding, and $i[2]$ is the number of new input reads after input forwarding. Therefore, dividing $i[2]$ by the number of channels mapped to one PE array is the stride for the selected $R$. The condition at line 9 checks if the previously calculated $st$ is in the integer domain and is less than the filter size ($R$). Then adversary can use Equation~\ref{eq:st_pd} to calculate $pd$ and check if $pd$ is in the integer domain and it is less than filter width ($R$). Line 14 calculates $Y'$ using the Equation \ref{eq:st_pd} from previously found values ($R,Y,st,pd$). The final condition asserts whether the side-channel information $O_w$ equals expected output writes ($X' \times Y' \times K$). The returned $H$ has potential solutions for the layer's parameters $\{R, K, C, st,pd\}$. In WS dataflow, $O_w$ is equal to the difference between the number of partial sum reads and writes ($psum_w - psum_r$).  Section~\ref{subsec:case_study_alex_ws} provides a case study on the second Conv layer of Alexnet, which converges into one solution.

\subsection{Recovery of Conv Layer from Output Stationary Dataflow}
\label{subsec:recover_conv_os}

Algorithm \ref{alg:os_attck_1} outlines the attack steps for recovering Conv layer parameters from the OS dataflow-supported architecture described in the Section~\ref{subsec:dataflow-based DNN accelerators}. Here we define the \textit{targeted event} for the cycle-to-cycle data collection as the cycle where the first \textit{ofmap} value is written. In other words, the adversary collects the cycle-to-cycle data from the first cycle to the cycle where the first \textit{output} is written. In an OS dataflow, there are no partial outputs/sums. In other words, an \textit{ofmap} value is written only after it is fully calculated. The intuition behind the \textit{targeted event} selection is to find the number of weight reads responsible for fully calculating one \textit{ofmap} value. 

\begin{algorithm}
\caption{Recovering Conv Layer from OS dataflow}
\label{alg:os_attck_1} 
\begin{algorithmic}[1] 
\Function{\textit{$recoverConv_{os}$}}{$W_r,O_w,w,i,o,X,Y,C$} 
\State $eq \leftarrow R^2C = t_e - 1$
\State $R \leftarrow$ solve $eq$ for $R$ in $\mathbb{Z}^+$
\State $st \leftarrow$ \&[$w_1$] - \&[$w_2$] 
\For{$pd=0$ to $R - 1$}
\State $X' \leftarrow$ substitute $R,st,pd,X$ for Equation~\ref{eq:st_pd}
\State $Y' \leftarrow$ substitute $R,st,pd,Y$ for Equation~\ref{eq:st_pd}
\If {$X', Y' \notin \mathbb{Z}^+$}

\textit{continue} \Comment{Not a solution}
\EndIf
\State $K \leftarrow$ substitute $\{X',Y'\}$ to Equation~\ref{eq:os_o_w}
\If {$K \in \mathbb{Z}^+$ and $\{X',Y',R,K,C\}$ satisfy Equation~\ref{eq:os_w_r}}
\State $H \leftarrow$ $H \cup $ $\{R,K,C,st,pd\}$
\EndIf
\EndFor
\State \textbf{return} $H$
\EndFunction
\end{algorithmic}
\end{algorithm}

When we consider one \textit{ofmap} value, it is generated from an accumulation of multiplications between a single filter of width and height of $R$ with $C$ channels. In other words, the total responsible weight reads for a single value in \textit{ifmap} is $R^2C$. Our selection of the \textit{targeted event} ensures that the ($t_e - 1$) equals $R^2C$ (line 2). Since the architecture reads one weight in a cycle, the relationship in line 2 holds. Solving this relationship in the $\mathbb{Z}^+$ domain gives the value of R. Because of the forwarding connections at each PE in the PE array, the difference between the virtual address of weight reads from the controller gives the value of stride (line 4). Lines 5 to 11 iterate through all possible $pd$ values. This loop termination considers that padding cannot exceed the filter width ($R$). Lines 6 and 7 calculate the width ($X'$) and height ($Y'$) of the \textit{ofmap} for the selected $pd$. The condition at line 8 checks if the calculated {$X', Y'$} is in $\mathbb{Z}^+$ domain; if not, we move to the next $pd$ value in the loop. In an OS dataflow, the number of total output written ($O_w$) to the GB:
\begin{equation} \label{eq:os_o_w}
O_{w} = X' \times Y' \times K 
\end{equation}

Line 9 substitutes previously calculated $\{X',Y'\}$ to the above equation and finds $K$. Line 9 checks for two conditions: (1) whether $K$ is in $\mathbb{Z}^+$ domain, and (2) does the relationship stated in Equation \ref{eq:os_w_r} holds for the potential value set $\{X',Y',R,K,C\}$?

\begin{equation} \label{eq:os_w_r}
W_{r} = (ceil(\frac{X'}{m}) \times ceil(\frac{Y'}{n}))R^2CK 
\end{equation}

All PEs in this OS architecture conducts MAC operations relevant to one output channel due to the broadcasting of weights. Therefore, the \textit{psum} accumulated at each registry is also relevant to a single output channel. When we zoom into Equation~\ref{eq:os_w_r}, $(ceil(\frac{X'}{m}) \times ceil(\frac{Y'}{n}))$ gives the number of tiles needed to calculate all \textit{ofmap} values of a single output channel. A tile is a subset/block of the output feature map that is processed as a unit. For example, if we process a layer with ($X', Y' = 12, K = 1 $) in an \textit{OS}($12,4$) architecture. The first tile calculates \textit{ofmap} values for the first four rows and the second and third tiles for the middle four rows and last four rows, respectively. From the first line of the algorithm, $R^2C$ gives weight reads for one tile. So, the number of tiles per single output channel $\times$ weight reads per tile $\times$ number of output channels gives the total weight reads. Section~\ref{subsec:case_study_alex_os} shows a case study on the second Conv layer of Alexnet that converged into one solution.
















\subsection{Extraction of FC Layer Parameters}
\label{subsubsec:extarcting_fc_layer}

An FC layer can be considered a Conv layer with a filter size equal to the size of the input, effectively connecting all neurons to each other. Therefore,  extracting FC layer parameters is relatively straightforward compared to extracting Conv layer parameters. An FC layer has only two parameters, which are input neuron size and output neuron size. FC layer parameters can be determined independently of the dataflow architecture by solely examining the total data reads and writes for the layer. The total number of input reads in a layer ($I_r$) equals the number of input neurons in an FC layer. Similarly, the total number of output writes ($O_w$) equals the number of output neurons. Additionally, a dense FC layer has $ W_r = I_r \times O_w $ relationship.

\subsection{Identification of Activation Functions}
\label{subsubsec:identification_of_activation}

Activation functions as integral components of a DNN's architecture. In our study, we assume that the accelerators incorporate the Rectified Linear Unit (ReLU) as the sole activation function, thereby eliminating the need for its explicit identification. The review of state-of-the-art dataflow CNN inference accelerators, including those detailed in \cite{chen2016eyeriss, chen2019eyeriss}, supports this assumption by consistently showing ReLU as sole implementation inside the accelerator. This choice is based on empirical evidence suggesting that CNNs typically achieve better performance with ReLU in their hidden layers~\cite{agarap2018deep}. The type of activation function employed in final layer can often be inferred based on the specific task that the model addresses. For instance, multi-class classification tasks generally use a softmax activation function, whereas binary classification tasks might use a sigmoid function. It is important to note that these final layer activations (e.g. sigmoid, softmax) are usually executed outside the dataflow accelerator, on the host CPU~\cite{sze2017efficient}. However if future dataflow accelerators support a variety of activation functions, such as Leaky ReLU or Parametric ReLU, our proposed method to identify the activation function is to monitor which specific functional unit is activated during the processing of each layer.

\subsection{Extraction of Pooling Layer Parameters}
\label{subsubsec:extractin_poool_layer}

Our methodology to extract pooling layers from max or average pooling depends on three assumptions based on typical pooling operations on CNNs. (1) Usually, the stride is greater than 1~\cite{sze2017efficient}. (2) Max and average pooling typically does not use padding. (3) CNNs tend to use small pooling filter sizes because large pooling filters tend to overfit models by losing  information.
The pooling operation does not change the number of output channels ($K$). Algorithm~\ref{alg:recover_pool} outlines the steps to recover pooling layer parameters ($\{R_{pool}, st_{pool}\}$) from side-channel information.

\begin{algorithm}
\caption{Recovering pooling layer}
\label{alg:recover_pool} 
\begin{algorithmic}[1] 
\Function{\textit{recoverPooling}}{$N_{pool}, X',Y',K$}
\For{$R_{pool}=2$ to $X'$}
\For{$st_{pool}=R_{pool}$ to $1$}
\State $X_{pool} \leftarrow ((X' - R_{pool})/st_{pool} + 1)$
\State $Y_{pool} \leftarrow ((Y' - R_{pool})/st_{pool} + 1)$
\If {$ N_{pool}/K = X_{pool} \times Y_{pool}$ }
\State \textbf{return} $\{R_{pool}, st_{pool}\}$
\EndIf
\EndFor
\EndFor
\EndFunction
\end{algorithmic}
\end{algorithm}

The loop at line 2 searches for pooling layers for increasing filter sizes, which gives more dominance to small pool layers. The second loop (line 3) ensures the attack first tries to match a non-overlapping pooling layer and increases the overlapping in subsequent iterations. Line 4 and 5 calculates the \textit{ofmap} width ($X_{pool}$) and height ($Y_{pool}$) after pooling. Finally, the condition at line 6 checks if the monitored $N_{pool}$ satisfies the calculated \{$X_{pool}, Y_{pool}$\} values to find the pooling layer parameters. If a particular CNN architecture employs a Global Average Pooling (GAP), we can use the same side-channel data ($N_{pool}$) to identify it. Unlike average and max pooling, which involve hyperparameters such as stride and filter size, GAP operates by computing the average of each feature map channel in the ofmap. This results in a single average value per channel, effectively producing a vector of size $K$ for an ofmap of dimensions $X' \times Y' \times K$. As GAP lacks configurable parameters beyond its basic function, the only information we can derive from side-channel analysis is the usage of GAP in a particular layer. If $N_{pool}$ corresponds exactly to the number of output channels ($K$), it indicates the use of GAP.


\section{Experiments}
\label{sec:experiments}

In this section, we first describe the experimental setup used to evaluate the effectiveness of our attack. Next, we present results on popular CNN models demonstrating potential structures recovered by our approach with two case studies on weight stationary and output stationary architectures. Then, we discuss the applicability and limitations of the proposed method. Finally, we outline potential mitigation techniques.

\subsection{Experimental Setup}
\label{subsec:experimental_setup}

We performed modifications on top of the cycle-accurate Stonne~\cite{munoz2021stonne} accelerator simulator to model the WS dataflow described in Section~\ref{subsec:ws_dataflow}. Furthermore, we extended Stonne, which originally only supported the WS dataflow, to also model the OS dataflow described in Section~\ref{subsec:os_dataflow}. We gathered  the side-channel information using the simulation stat of the global buffer of the simulator. We provided adequate bandwidth for interconnects of each data type (input, output, weight) in each simulation so that any data loads will happen in one cycle (Section~\ref{subsec:applicability} discuss relaxing of this assumption). We modeled a simple pooling module in both accelerators to mimic pooling operations and extract the number of pooling operations. We modeled three concrete weight stationary accelerator architectures from the architecture described in Section~\ref{subsec:dataflow-based DNN accelerators}: \textit{WS}($4,4$), \textit{WS}($12,4$), and \textit{WS}($24,10$). Similarly, we used three output stationary dataflow accelerators: \textit{OS}$(4,4)$, \textit{OS}$(10,4)$, and \textit{OS}$(20,10)$. 
We attacked and recovered popular CNN models: a 5-layer Lenet, an 8-layer AlexNet, a 16-layer VGGnet-16, and 28-layer YOLOv2 to evaluate the proposed CNN model recovery attack. Each CNN model was subjected to 10 rounds of experiments to ensure robustness and repeatability of results. Table~\ref{tab:final-results} shows an overview of composition of the layers in these models.  

We conduct experiments on real hardware implementations of the two weight stationary accelerator architectures (\textit{WS(4,4), and WS(12,4)}). These architectures were implemented on the Alchitry Cu FPGA board, which utilizes the Lattice iCE40 HX FPGA supported by the open source tool chain IceStorm. The configuration parameters of the accelerator implementation in our experiments are outlined in Table~\ref{tab:acc-config}.

\begin{table}[htp]
\centering
\caption{WS accelerator configuration parameters}
\vspace{-0.1in}
\label{tab:acc-config}
\resizebox{0.50\textwidth}{!}{%
\begin{tabular}{|l|l|}
\hline
\textbf{Parameter} & \textbf{Value} \\ \hline
PE array dimensions: (m,n) & {[}(4,4), (12,4){]} \\ \hline
Precision (weights, ifmap, ofmap) & 8 bits \\ \hline
GB to PE arrays input bus width & $4 \times 8$ bits \\ \hline
GB to PE arrays weight bus width & $4 \times 8$ bits \\ \hline
Accumulator tree precision & 32 bits \\ \hline
Accumulated output bus width & $4 \times 32$ bits \\ \hline
Input forwading path width & 8 bits \\ \hline
Global buffer size & 256B \\ \hline
\end{tabular}%
}
\end{table}

Due to space limitations on the FPGA, we opted for a relatively small global buffer (GB). However, this constraint does not affect our ability to accurately recover side-channel information regarding the number of memory reads and writes from GB. We use a Python frontend, similar to the one used in STONNE~\cite{munoz2021stonne}, to emulate the host CPU. The UART (Universal Asynchronous Receiver/Transmitter) communication protocol is employed to facilitate serial communication between the Python frontend and the FPGA-implemented accelerator. This setup allows us to send control and data to the accelerator and receive snooped side-channel information from the accelerator. To facilitate the capture of memory-based side-channel information, particularly the number of memory reads and writes from the GB, we used snooping on the bus between the GB and the processing element array (\redcircleletter{a}
 in Figure \ref{fig:overview_of_accel}), as well as between the DRAM and GB (\redcircleletter{c}
 in Figure \ref{fig:overview_of_accel}). Snooping of buses was achieved by embedding two Hardware Trojans (HTs) into the RTL of the accelerator, analogous to the use of hardware performance counters in GPUs to emulate snooping, as discussed in previous studies~\cite{hu2020deepsniffer}. We conducted 10 rounds of experiments for each of the four CNN models: Lenet, AlexNet, VGGnet-16, and YOLOv2. 

\begin{table}[tp]
\centering
\caption{Number of potential structures recovered from Lenet, Alexnet, VGGnet-16 and YOLOv2 using our method and comparison with ~\cite{hua2018reverse} that exploit DRAM to GB memory access on temporal inference accelerator.}
\resizebox{0.60\textwidth}{!}{%
\begin{tabular}{|ll|l|c|c|c|}
\hline
\multicolumn{2}{|l|}{\textbf{CNN model}} & \textbf{Lenet} & \textbf{Alexnet} & \textbf{VGGnet-16} & \multicolumn{1}{l|}{\textbf{YOLOv2}} \\ \hline
\multicolumn{2}{|l|}{\textbf{\begin{tabular}[c]{@{}l@{}}Number of layers\\ (Conv/Pool/FC)\end{tabular}}} & 3/2/2 & 5/3/3 & 13/5/3 & 23/5/0 \\ \hline
\multicolumn{2}{|l|}{\textbf{Conv layer filter sizes}} & \begin{tabular}[c]{@{}l@{}}5x5\\ 2x2\end{tabular} & \begin{tabular}[c]{@{}c@{}}11x11\\ 5x5, 3x3\end{tabular} & \begin{tabular}[c]{@{}c@{}}1x1\\ 3x3\end{tabular} & \begin{tabular}[c]{@{}c@{}}1x1\\ 3x3\end{tabular} \\ \hline
\multicolumn{1}{|l|}{\multirow{7}{*}{\textbf{\begin{tabular}[c]{@{}l@{}}Number of \\ potential\\ structures\end{tabular}}}} & \textit{WS(4,4)} & 1 & 1 & 1 & 1 \\ \cline{2-6} 
\multicolumn{1}{|l|}{} & \textit{WS(12,4)} & 1 & 1 & 1 & 1 \\ \cline{2-6} 
\multicolumn{1}{|l|}{} & \textit{WS(24,10)} & 1 & 1 & 1 & 1 \\ \cline{2-6} 
\multicolumn{1}{|l|}{} & \textit{OS(4,4)} & 1 & 1 & 1 & 1 \\ \cline{2-6} 
\multicolumn{1}{|l|}{} & \textit{OS(10,4)} & 1 & 1 & 1 & 1 \\ \cline{2-6} 
\multicolumn{1}{|l|}{} & \textit{OS(20,10)} & 1 & 1 & 1 & 1 \\ \cline{2-6} 
\multicolumn{1}{|l|}{} & \cite{hua2018reverse} & 9 & 24 & - & - \\ \hline
\end{tabular}
}
\label{tab:final-results}
\end{table}

\subsection{Results}

As shown in Table~\ref{tab:final-results}, we can fully recover CNN parameters for Lenet, Alexnet, VGGnet-16, and YOLOv2 for both WS and OS dataflow accelerators of three sizes, across all ten iterations per a model. Furthermore, it shows our approach can recover exact model architectures compared to model recovery via snooping DRAM to GB memory access~\cite{hua2018reverse} on temporal inference accelerator that does not consider a dataflow. The layer boundaries of WS dataflow are identified by monitoring the configuration phase and OS dataflow by observing RAW dependency of feature maps. The CNN structure has converged into one solution, which highlights that for each Conv layers Algorithm \ref{alg:ws_attck_1} and \ref{alg:os_attck_1} converged to one solution ($size(H)=1$) for the respective layer parameters. This is because our method can recover exact values for some parameters ($X'$ in WS and $R$ and $st$ in OS) and use them with multiple condition checks to converge to the potential structure. Our findings from these real-world experiments on FPGA implemented accelerator align with the results from cycle-accurate simulation, successfully recovering the exact architecture of all four CNN models every time, across all tested WS dataflow configurations.

\begin{table}[htp]
\centering
\caption{Side channel data ($I_r, O_w, W_r$) for all fully connected layers of Alexnet on OS and WS dataflows.}
\resizebox{0.45\textwidth}{!}{%
\begin{tabular}{|l|l|l|l|}
\hline
Layer & FC1 & FC2 & FC3 \\ \hline
Input reads & 9216 & 4096 & 4096 \\ \hline
Weight reads & 37748736 & 16777216 & 4096000 \\ \hline
Output writes & 4096 & 4096 & 1000 \\ \hline
\end{tabular}%
}
\label{tab:sc_fc_alexnet}
\end{table}

Since no approaches exist on dataflow architectures to recover CNN models, we compare our results with \cite{hua2018reverse}, which uses the DRAM to GB memory access patterns and execution time of layers on a temporal accelerator to recover CNN models. They were able to recover six potential structures for Lenet and 24 potential structures for Alexnet. Since our approach exploits inherent dataflow patterns and data reuse that leak critical characteristics of layers, our solution converged into one correct structure in both cases across all architectures. This accurate recovery of CNN architecture, facilitated by architectural hints from dataflow inference accelerators, highlights how layer-specific mapping and processing enhance the leakage of CNN architectures. Table \ref{tab:sc_fc_alexnet} and \ref{tab:sc_pool_alexnet} shows side-channel data on all FC and pooling layers of Alexnet. The side-channel data of FC and pooling layers do not depend on the specific dataflow architecture of the accelerator. Section \ref{subsec:case_study_alex_ws} and \ref{subsec:case_study_alex_os} provide two case studies using Alexnet to provide insight into the recovery procedure of Conv layers.

\begin{table}[htp]
\centering
\caption{Side channel data ($N_{pool}$) for all pooling layers of Alexnet running on the pooling module.}
\resizebox{0.45\textwidth}{!}{%
\begin{tabular}{|l|l|l|l|}
\hline
Layer & Pool 1 & Pool 2 & Pool 3 \\ \hline
Number of pool op. & 69984 & 43264 & 9216 \\ \hline
\end{tabular}%
}
\label{tab:sc_pool_alexnet}
\end{table}


\subsection{Case Study: Alexnet with Weight Stationary Dataflow}
\label{subsec:case_study_alex_ws}


Table \ref{tab:sc_ws_alexnet} shows the side-channel information used to recover all five Conv layers of Alexnet in the \textit{WS}($12,4$) accelerator. The attack needs only a relatively small number of cycle-wise data to be collected. For example, the \textit{targeted event} cycle in the second Conv layer is 27, while the total is 11197441 cycles. The rest of the section zooms in on recovering the second Conv layer of Alexnet using Algorithm~\ref{alg:ws_attck_1}.

\begin{table}[htp]
\centering
\caption{Side-channel values for five Conv layers of Alexnet running on \textit{WS}($12,4$) and the number of cycles.}
\resizebox{0.55\textwidth}{!}{%
\begin{tabular}{|l|l|l|l|l|l|}
\hline
Layer & \begin{tabular}[c]{@{}l@{}}Weight \\ reads\end{tabular} & \begin{tabular}[c]{@{}l@{}}\textit{psum} \\ reads\end{tabular} & \begin{tabular}[c]{@{}l@{}}\textit{psum} \\ writes\end{tabular} & \begin{tabular}[c]{@{}l@{}}Event\\ cycle\end{tabular} & \begin{tabular}[c]{@{}l@{}}Total\\ cycles\end{tabular} \\ \hline
Conv1 & 34848 & 9292800 & 9583200 & 55 & 2395801 \\ \hline
Conv2 & 614400 & 44603136 & 44789760 & 27 & 11197441 \\ \hline
Conv3 & 884736 & 12395136 & 12460032 & 13 & 3115009 \\ \hline
Conv4 & 1327104 & 18625152 & 18690048 & 13 & 4672513 \\ \hline
Conv5 & 884736 & 12416768 & 12460032 & 13 & 3115009 \\  \hline
\end{tabular}
}
\label{tab:sc_ws_alexnet}
\end{table}

\begin{figure*}[htp]
\centering
\includegraphics[width=0.90\textwidth]{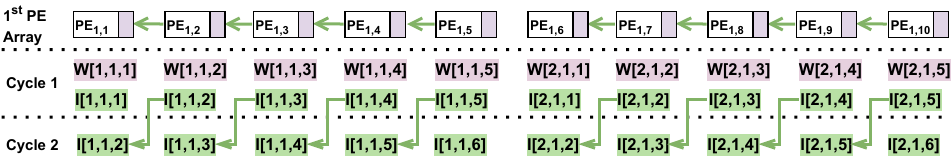}
\caption{Dataflow mapping at 1$^{st}$ and 2$^{nd}$ cycle of Alexnet Conv2 layer in \textit{WS}(12,4) accelerator (only 1$^{st}$ PE array is shown).}
\label{ffig:example_ws_alex_conv2_mapping}
\end{figure*}

The second Conv layer of Alexnet has 256 filters ($K =256$) with parameters $\{R=5, C=96, st=1, pd=2\}$. This layer takes a input feature map of size $\{X=27, Y=27, C=96\}$ and output a feature map of paramaters $\{X'=27, Y'=27, K=256\}$. According to the dataflow mapping described in Section \ref{subsec:dataflow-based DNN accelerators}, this layer is mapped as two rows (row size = 5) of filter representing two input channels in one PE array. There are four such PE arrays representing different filters. Figure \ref{ffig:example_ws_alex_conv2_mapping} shows the mapping of dataflow in the first two cycles in the first PE array with ten active PEs (the diagram does not show unmapped and idle PE(1,11) and PE(1,12)). The mapping of weights maximizes the input forwarding between two consecutive cycles (there are only two new input reads and eight forwarding in second cycle). The other three PE arrays load weights of filter 2-4 in the same relative order and use the same \textit{ifmap} values provided to the first PE array through array-wise multicast.

\begin{figure}[htp]
\includegraphics[width=0.65\textwidth]{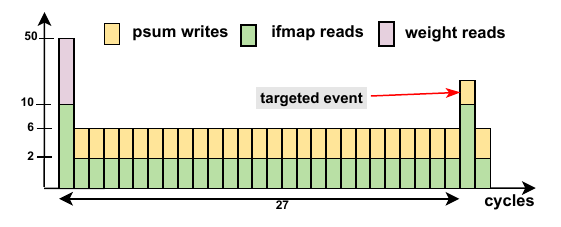}
\vspace{-0.1in}
\caption{Cycle-wise number of data reads/writes until the \textit{targeted event} of Alexnet Conv2 layer in \textit{WS}(12,4).}
\label{fig:example_ws_alex_conv2_graph}
\end{figure}

Execution of the Conv2 of Alexnet in \textit{WS}($12,4$) results in $W_r = 614400$, $psum_r = 44603136$, and $psum_w = 44789760$. The rest of the section goes through the Algorithm \ref{alg:ws_attck_1} to recover Conv2 parameters of Alexnet. Figure~\ref{fig:example_ws_alex_conv2_graph} shows the cycle-by-cycle number of GB reads/writes of different data types. The set H with potential (R,K) values = \{(2,1600), (4,400), (5,256), (8,100), (10,64), (16,25), (20,16), (40,4), (80,1)\}. Then the number of active PE arrays can be calculated as $n_a = 40/10 = 4$. Since the \textit{targeted event} occurs in the 27$^{th}$ cycle, $X'$ is 27. Then let us look at each condition and what potential values are filtered out. The first condition (line 6) applies to the first five elements of $H$ where $R \leq 10$. From these five values, (4,400) and  (8,100) are filtered out from $H$. For example $(w^1/n_a)\%R$ is $10\%4 = 2$ for (4,400).

When considering the next two conditions in lines 9 and 12, Table \ref{tab:example_conv2_ws} shows $st$ and $pd$ values generated according to the algorithm. $i[2]=2$ is from side channels, which is used to calculate $st$ for each value remaining in set $H$. When we look at the table, only (5,256) satisfy both conditions. It is important to notice that potential $R \geq 10$ fails the $pd > R$ condition. When we consider the final condition and the remaining value of set H (5,256): $X' \times Y' \times K = 186624$, which is equal to $O_w = psum_w - psum_r$. Therefore, we can successfully recover Alexnets' Conv2 layer parameters. 


\begin{table}[htp]
\centering
\caption{Checking conditions 2 and 3 of Algorithm~\ref{alg:ws_attck_1} for Conv2 of Alexnet on \textit{WS}($12,4$) dataflow accelerator.}
\label{tab:example_conv2_ws}
\resizebox{0.7\textwidth}{!}{%
\begin{tabular}{|l|c|c|c|c|c|c|c|}
\hline
\textbf{(R,K)} & \multicolumn{1}{l|}{(2,1600)} & \multicolumn{1}{l|}{(5,256)} & \multicolumn{1}{l|}{(10,64)} & \multicolumn{1}{l|}{(16,25)} & \multicolumn{1}{l|}{(20,16)} & \multicolumn{1}{l|}{(40,4)} & \multicolumn{1}{l|}{(80,1)} \\ \hline
\textbf{max(m//R,1)} & 6 & 2 & 1 & 1 & 1 & 1 & 1 \\ \hline
\textbf{st} & 4/6 & 1 & 2 & 2 & 2 & 2 & 2 \\ \hline
\textbf{pd} & - & 2 & 35/2 & 41/2 & 45/2 & 65/2 & 105/2 \\ \hline
\textbf{x/\checkmark} & x & \checkmark & x & x & x & x & x \\ \hline
\end{tabular}
}
\end{table}

\subsection{Case Study: Alexnet with Output Stationary Dataflow}
\label{subsec:case_study_alex_os}


Table \ref{tab:sc_os_alexnet} shows the side-channel information used to recover all the five convolution layers of Alexnet in the \textit{OS}($10,4$) accelerator. The attack needs only a small number of cycle-wise data to be collected. For example, the \textit{targeted event} cycle in the first Conv layer is 363, while the total is 2927233 cycles. The remainder of the section describes how to recover the first Conv layer of Alexnet using Algorithm~\ref{alg:os_attck_1}.

\begin{table}[htp]
\centering
\caption{Side-channel information for all Conv layers on Alexnet running on \textit{OS}($10,4$)}
\vspace{-0.05in}
\resizebox{0.45\textwidth}{!}{%
\begin{tabular}{|l|l|l|l|l|}
\hline
Layer & \begin{tabular}[c]{@{}l@{}}Weight \\ reads\end{tabular} & \begin{tabular}[c]{@{}l@{}}Output \\ writes\end{tabular} & \begin{tabular}[c]{@{}l@{}}Event\\ cycle\end{tabular} & \begin{tabular}[c]{@{}l@{}}Total\\ cycles\end{tabular} \\ \hline
Conv1 & 2927232 &  290400  & 363 & 2927233 \\ \hline
Conv2 & 12902400  &  186624  & 2400 & 12902401 \\ \hline
Conv3 & 7077888 &  64896  & 2304 & 7077889 \\ \hline
Conv4 & 10616832 &  64896  & 3456 & 10616833 \\ \hline
Conv5 & 7077888 &  43264  & 3456 & 7077889 \\ \hline
\end{tabular}
}
\label{tab:sc_os_alexnet}
\end{table}

\begin{figure*}[htp]
\centering
\includegraphics[width=0.90\textwidth]{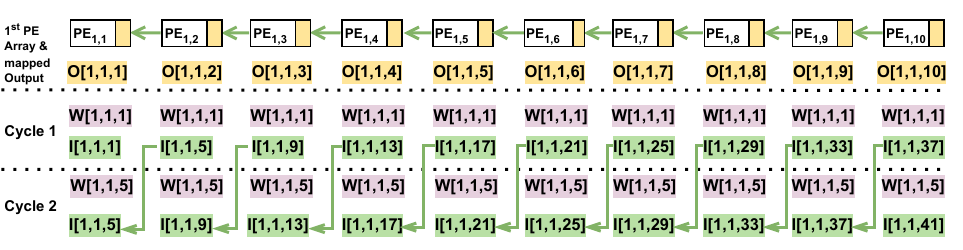}
\vspace{-0.1in}
\caption{Dataflow mapping at 1$^{st}$ and 2$^{nd}$ cycle of Alexnet Conv1 layer in \textit{OS}($10,4$) accelerator (only 1$^{st}$ PE array is shown).}
\label{fig:example_os_alex_conv1_mapping}
\vspace{-0.10in}
\end{figure*}

The first Conv layer of Alexnet has 96 filters ($K =96$) with parameters $\{R=11, C=3, st=4, pd=0\}$. This layer takes an input feature map of size $\{X=227, Y=227, C=3\}$ and outputs a feature map of parameters $\{X'=55, Y'=55, K=96\}$. Figure \ref{fig:example_os_alex_conv1_mapping} shows dataflow mapping in the first two cycles: the first ten entries of the first row of the \textit{ofmap} are accumulated in the first row of the PE array. Similarly, the subsequent three rows of the \textit{ofmap} are mapped into the next three PE arrays in order. The weight reads in consecutive cycles are done to maximize input forwarding. For example, the reading of the $W[1,1,5]$ in the second cycle after $W[1,1,1]$ in the first cycle results in only one input read from GB ($I[1,1,41]$) for that PE array. Only four input reads from GB in the second cycle for all four PE arrays.

\begin{figure}[htp]
\includegraphics[width=0.7\textwidth]{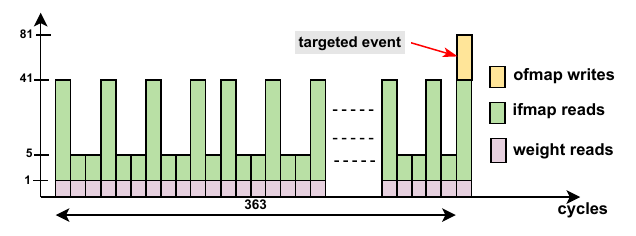}
\vspace{-0.2in}
\caption{Cycle-wise number of data reads/writes until the \textit{targeted event} of Alexnet Conv1 layer in \textit{OS}(10,4).}
\label{fig:example_os_alex_conv1_graph}
\end{figure}

Execution of the Conv1 of Alexnet in \textit{OS}($10,4$) results in $W_r = 2927232$ and $O_w = 290400$. The rest of the section goes through the Algorithm \ref{alg:os_attck_1} to recover Conv1 parameters of Alexnet. Figure \ref{fig:example_os_alex_conv1_graph} shows the cycle-by-cycle number of GB reads/writes of different data types. The cycle of the \textit{targeted event} ($t_e$) is 364. Therefore, $R^2C = 363$ (line 1). Solving this in the $\mathbb{Z}^+$ domain gives $R=11$. The virtual address difference between weight reads in the first (W[1,1,5]) and second (W[1,1,1]) cycle is 4, which is equal to the stride. When we consider the two conditions in lines 8 and 10, Table \ref{tab:example_conv1_os} shows $X'/Y'$ and $K$ values generated according to the algorithm for each potential $pd$ value (0-10). As shown in the table, only $pd=0$ passes all the conditions. $pd =0$ also satisfies Equation \ref{eq:os_w_r} ($2927232 = (ceil(55/10) \times ceil(55/4))11^2 \times 3 \times 96$ )). Therefore, we can successfully recover Alexnets' Conv1 layer parameters.

\begin{table}[htp]
\centering
\caption{Checking conditions 1 and 2 of Algorithm \ref{alg:os_attck_1} for Conv1 of Alexnet on \textit{OS}($10,4$) dataflow accelerator.}
\vspace{-0.05in}
\resizebox{0.7\textwidth}{!}{%
\begin{tabular}{|l|c|c|c|c|c|c|c|c|c|c|c|}
\hline
\textbf{pd} & 0 & 1 & 2 & 3 & 4 & 5 & 6 & 7 & 8 & 9 & 10 \\ \hline
\textbf{X'/Y'} & 55 & 55.5 & 56 & 56.5 & 57 & 57.5 & 58 & 58.5 & 59 & 59.5 & 60 \\ \hline
\textbf{K} & 96 & - & 92.6 & - & 89.3 & - & 86.3 & - & 83.4 & - & 80.6 \\ \hline
\textbf{x/\checkmark} & \checkmark & x & x & x & x & x & x & x & x & x & x \\ \hline
\end{tabular}%
}
\vspace{-0.1in}
\label{tab:example_conv1_os}
\end{table}

\subsection{Applicability and Limitations}
\label{subsec:applicability}

Our proposed attack works on any concrete configuration of abstract WS and OS dataflow architectures outlined in Section \ref{sec:background}. For this study, we assumed there is no bandwidth limitation on interconnect for each datatype. After relaxing this assumption, a minor modification in Algorithm \ref{alg:ws_attck_1} in WS dataflow or \ref{alg:os_attck_1} in OS dataflow can generate the same results. For example, if weight NoC has a bandwidth limitation of 10 on the processing of the Conv layer in \textit{WS}($12,4$), as elaborated in Figure \ref{fig:example_ws_alex_conv2_graph}, four cycles would be spent on the initial weight memory read. We can identify the initial four memory reads by either lagging of \textit{psum} writes by three cycles or idling input interconnect for three cycles. Another assumption we made was the accumulation of multiplication on one PE array in WS dataflow happens in one cycle. If we relax this assumption and set \textit{q} cycles for accumulating a \textit{psum} write in a PE array, every \textit{psum} write will lag by extra $q-1$ cycles. A minor modification of calculating the event cycle as $t_e - q$ can fix this. Our approach can be applied to folding-supported WS architectures since our attack uses side-channel information independent of folding. 

One feature of our methodology warrants further discussion, is its approach to scenarios where multiple potential structures for a previous layer might exist. In such cases, our algorithm tests the ifmap parameters of each potential structure to identify the correct configuration for the subsequent layer. Our approach utilizes a layer-by-layer structure extraction mechanism where integer factorization establishes an initial set of possible outcomes. These are then refined through a set of rules driven by architectural hints and side-channel information, ensuring convergence to a singular structure for each layer. Our experimental results, covering 64 layers across six different dataflow accelerators (384 combinations), consistently demonstrate a convergence to a single layer structure, proving the robustness of our approach. Furthermore, if multiple structures exist in the previous layer, the incompatibility with rules in the current or any preceding layer will further narrow down the solutions. Due to the use of a layer-by-layer extraction approach, our method is inherently scalable for more complex and deeper architectures; this is evidenced by its successful scaling for deeper networks in our results. 

In this paper, we develop a methodology for recovering CNN models from weight stationary and output stationary dataflow architectures with input forwarding. The proposed attack can be extended to any WS accelerator with input forwarding, such as MAERI~\cite{kwon2018maeri}, by leveraging the hierarchical interconnects and input forwarding mechanisms to infer memory access patterns and extract CNN model parameters. The main difference between MAERI and the WS architecture in the paper lies in their reconfigurable interconnects, allowing dynamic adjustment of PE array size when processing each layer. This control message for reconfiguration can be snooped from the same bus (\redcircleletter{d} in Figure \ref{fig:overview_of_accel}) used in layer boundary identification (Section~\ref{subsec:identify_layer_boundary}). Once the PE size is determined through snooping, the proposed attack can be applied accordingly.
Our attack on OS dataflow can be extended to other OS architectures with input forwarding ~\cite{du2015shidiannao,moons201714} to recover CNN models. Both  architectures, ShiDianNao~\cite{du2015shidiannao} and Envision~\cite{moons201714}, adhere to the OS architectures with input forwarding outlined in the paper, with subtle implementation differences. For example, Envision, rather than having dedicated input forwarding at each PE array, uses a FIFO buffer shared between PE arrays for input forwarding. Since this does not change the observed side-channel information (number of reads and writes) outside of the PE arrays, our attack can be extended to this architecture as well. These extensions are feasible because both WS and OS architectures share fundamental characteristics in data movement and reuse that were exploited in the attacks to recover model parameters and structures. The underlying abstract methodology can be extended to recover CNN models from other dataflow architectures using different local forwarding data-type (e.g., Neuflow~\cite{farabet2011neuflow}) and different dataflow types (input-stationary and row stationary~\cite{chen2016eyeriss}) by (1) defining a \textit{targeted event}/\textit{events} to collect cycle-wise number of data reads and writes, (2) collecting total reads and writes responsible for a layer, and (3) exploiting spatial and temporal data reuse in (1) and (2) with architectural details. Our approach cannot be directly applied to recover CNN structures with sparse FC and Conv layers such as Squeezenet~\cite{iandola2016squeezenet}. Our study on recovering CNN architectures highlights that memory access patterns should not be exposed to adversaries to avoid the leaking of CNN model architectures through dataflow-based CNN accelerators.

\subsection{Potential Mitigation Techniques}
\label{subsec:potential_mitigation_techniques}

We highlight potential countermeasures that can be explored by designers in securing CNN architectures against proposed memory-based side channel attacks. Several mitigation strategies can be explored by the designers. (1) Introducing dummy operations that do not affect the final output but alter the memory access patterns and computational behavior can help obscure the true operations being executed, thereby confusing attackers and complicating their analysis of side-channel data. (2) Implementing variable precision for weights and activations can make it more challenging for attackers to determine the number of memory reads and writes, increasing the difficulty of correlating observed memory operations with specific network layer structures. (3) Employing decoy architectures and alternative data paths that activate during specific operations can mislead attackers about the actual computational processes. (4) Specifically for OS dataflow architectures, randomizing the order of data storage in memory, such as weights and activations, can prevent attackers from easily correlating observed memory access patterns with specific operations. These countermeasures can also be combined to form a multi-layered defense strategy against side-channel attacks. Future research could further explore these mitigation techniques to validate their effectiveness against model recovery attacks in dataflow-based inference accelerators. 
\section{conclusion}
\label{sec:conclusion}

Artificial intelligence at edge devices is becoming increasingly ubiquitous with the abundance of data. Convolution neural networks (CNN) are executed using dataflow-based CNN accelerators due to energy efficiency. These accelerators use dataflows coupled with architectural designs to maximize different types of data reuse in CNN layers to efficiently perform inference using CNN models. This paper proposes an end-to-end memory-based side-channel attack that exploits dataflow patterns with the help of architectural hints to recover CNN model structures. Extensive evaluation of multiple architectures on weight stationary and output stationary dataflows demonstrates that our proposed method can fully recover well-known benchmark CNN models running in these CNN accelerators. This work also highlights the importance of concealing memory access patterns in dataflow-based inference accelerators. 

\section*{Acknowledgments}
This work was partially supported by National Science Foundation (NSF) grant SaTC-1936040.


\bibliographystyle{ACM-Reference-Format}
\bibliography{bibliography}


\end{document}